%% file: main.tex
\documentclass[sigconf]{nonacmart}
\setcopyright{rightsretained}
\acmDOI{10.18122/B2GQ53}

\acmISBN{}

\setcopyright{rightsretained}

\acmConference[RMSE'19]{ACM RMSE Workshop}{September 2019}{Copenhagen, Denmark} 
\acmYear{2019}
\copyrightyear{2019}

\acmPrice{15.00}

\usepackage{graphicx}  %Required
\usepackage{booktabs} % For formal tables
\usepackage{caption}
\usepackage{subcaption}
\usepackage{enumerate}
\usepackage{multirow}
\newcommand{\algname}[1] {{\fontfamily{cmtt}\selectfont {#1}}}

\renewcommand\footnotetextcopyrightpermission[1]{}

\begin{document}

\title[Bias Disparity in Collaborative Recommendation]{Bias Disparity in Collaborative Recommendation: Algorithmic Evaluation and Comparison}
\titlenote{Copyright 2019 for this paper by its authors. Use permitted under Creative Commons License Attribution 4.0 International (CC BY 4.0).\\Presented at the RMSE workshop held in conjunction with the 13th ACM Conference on Recommender Systems (RecSys), 2019, in Copenhagen, Denmark.}

\author{Masoud Mansoury}
\authornote{This author also has affiliation in School of Computing, DePaul University, Chicago, USA, mmansou4@depaul.edu.}
\affiliation{%
  \institution{Eindhoven University of Technology}
  %\streetaddress{P.O. Box 1212}
  \city{Eindhoven}
  \state{the Netherlands}
}
\email{m.mansoury@tue.nl}

\author{Bamshad Mobasher}
\affiliation{%
  \institution{DePaul University}
  %\streetaddress{1 Th{\o}rv{\"a}ld Circle}
  \city{Chicago}
  \country{USA}}
\email{mobasher@cs.depaul.edu}

\author{Robin Burke}
\affiliation{%
  \institution{University of Colorado Boulder}
  \city{Boulder}
  \country{USA}}
\email{robin.burke@colorado.edu}

\author{Mykola Pechenizkiy}
\affiliation{%
 \institution{Eindhoven University of Technology}
 %\streetaddress{Rono-Hills}
 \city{Eindhoven}
 %\state{Arunachal Pradesh}
 \country{the Netherlands}}
\email{m.pechenizkiy@tue.nl}

\renewcommand{\shortauthors}{Masoud Mansoury, et al.}

\begin{abstract}
  Research on fairness in machine learning has been recently extended to recommender systems. One of the factors that may impact fairness is bias disparity, the degree to which a group's preferences on various item categories fail to be reflected in the recommendations they receive. In some cases biases in the original data may be amplified or reversed by the underlying recommendation algorithm. In this paper, we explore how different recommendation algorithms reflect the tradeoff between ranking quality and bias disparity. Our experiments include neighborhood-based, model-based, and trust-aware recommendation algorithms.
\end{abstract}

%\begin{CCSXML}
%<ccs2012>
%  <concept_id>10010520.10010553.10010562</concept_id>
%  <concept_desc>Human-centered computing~Collaborative and social computing</concept_desc>
%  <concept_significance>500</concept_significance>
%  <concept_id>10010520.10010575.10010755</concept_id>
%  <concept_desc>Human-centered computing~Social recommendation</concept_desc>
%  <concept_significance>300</concept_significance>
%  <concept_id>10003033.10003083.10003095</concept_id>
%  <concept_desc>Information Retrieval~Retrieval efficiency</concept_desc>
%  <concept_significance>100</concept_significance>
%</ccs2012>
%\end{CCSXML}

%\ccsdesc[500]{Human-centered computing~Collaborative and social computing}
%\ccsdesc[300]{Human-centered computing~Social recommendation}
%\ccsdesc[100]{Information Retrieval~Retrieval efficiency}

\keywords{Recommender systems, Trust ratings, Fairness, Bias disparity}

\maketitle

\section{Introduction}
%{\let\thefootnote\relax\footnote{  Workshop on Recommendation in Multi-Stakeholder Environments (RMSE) at ACM RecSys 2019, Copenhagen, Denmark.}}
Recommender systems are powerful tools in extracting users preferences and suggesting desired items. These systems, while accurate, may suffer from a lack of fairness to specific groups of users. Research in fairness-aware recommender systems have shown that the outputs of recommendation algorithms are, in some cases, biased against protected groups \cite{ekstrand2018}. As a result, this discrimination among users will degrade users' satisfaction, loyalty, and effectiveness of recommender systems, and at worst, it can lead to or perpetuate undesirable social dynamics. 

Discrimination in recommendation output can originate from different sources. It may stem from the underlying biases in the input data \cite{virginia2018,burke2017} used for training. On the other hand, the discriminative behavior may be the result of recommendation algorithms \cite{kamishima2011,zembel2013,yao2017}.  

In this paper, we examine the effectiveness of recommendation algorithms in capturing different groups' interests across item categories. We compare different recommendation algorithms in terms of how they capture the categorical preferences of users and reflect them in the recommendation delivered. 
%of how categories of recommended items for a group of users, while accurate, deviates from a group's normal preferences. 

%We consider this analysis as consumer-side fairness as we are interested to low disparity in recommended item categories for each group of users. 

%Besides consumer-side fairness, we also take into account item coverage in recommended lists as provider-side fairness. 

It is important to note that in this paper, although we do not directly measure the fairness of recommendation algorithms, we study bias disparity of recommendation algorithms as an important factor that affects fairness. The benefit of studying bias disparity in recommender systems is that, depending on the domain, knowing which algorithms produce more or less disparity from users' stated preferences can allow system designers to better control the recommendation output. In our analysis of bias disparity, we also take into account item coverage in recommended lists. A recommendation algorithm with higher item coverage signifies that majority of item providers in the system will have equal chance to be shown to users. 

Our analysis includes a variety of recommendation algorithms: neighborhood models, factorization models, and trust-aware recommendation algorithms. In particular we investigate the performance of trust-aware recommendation algorithms. In these algorithms, besides items ratings, explicit trust ratings are used as side information to enhance the quality of input values for recommender systems. It has been shown that using explicit trust ratings will provide advantages for recommender systems \cite{Massa:2007a}. First, since trust ratings can be propagated, they can help overcome cold-start issue in recommender systems. Secondly, trust-aware methods are robust against shilling attacks in recommender systems \cite{Lam:2004a}. In this paper, we also analyze the performance of these algorithms in addressing bias disparity in recommender systems. 

The motivation behind this research is analyzing the performance of recommendation algorithms in preference deviation across item categories for a specific group of users (e.g., male vs. female). Given protected and unprotected groups, we aim to compare the ability of recommendation algorithms to generate recommendations equally well for each group based on their preferences in training data. Therefore, no matter what the context of the dataset is, given protected/unprotected groups and item categories, we are interested in comparing recommendation algorithms for their ability to recommend preferred item categories to these groups of users.      

For experiments, we prepared a sample of publicly-available Yelp dataset for research on fairness-aware recommender systems. Our experiments are performed on multiple recommendation algorithms and the results are evaluated in terms of \textit{bias disparity} and \textit{average disparity} along with ranking quality and item coverage. 

\section{Background}

The problem of unfair outputs in machine learning applications is well studied \cite{kamiran2010, dwork2012, bozdag:2013} and also it has been extended to recommender systems. Various studies have considered fairness in recommendation results \cite{burke2017}.  

%\subsection{Consumer Fairness}

One research direction in fairness-aware recommender systems is providing fair recommendations for consumers. Burke et. al. in \cite{burke2017} have shown that inclusion of a balanced neighborhood regularization to SLIM algorithm can improve the equity of recommendations for protected and unprotected groups. Based on their definition for protected and unprotected groups, their solution takes into account the group fairness of recommendation outputs. Analogously, Yao and Huang in \cite{yao2017} improved the equity of recommendation results by adding fairness terms to objective function in model-based recommendation algorithms. They proposed four fairness metrics that capture the degree of unfairness in recommendation outputs and added these metrics to learning objective function to further optimize it for fair results. 

Zhu et al. in \cite{zhu2018} proposed a fairness-aware tensor-based recommender systems to improve the equity of recommendations while maintaining the recommendation quality. The idea in their paper is isolating sensitive information from latent factor matrices of the tensor model and then using this information to generate fairness-aware recommendations.

%\subsection{Provider Fairness}

Besides consumer fairness, provider fairness is another research direction in fairness-aware recommender systems. Provider fairness refers to the fact that items belong to each provider have equal chance to be shown in the recommended lists. This is known as \textit{popularity bias} and usually measured by \textit{item coverage}. 

Abdollahpouri et al., \cite{himan2017} addressed popularity bias in learning-to-rank algorithms by inclusion of fairness-aware regularization term into objective function. They showed that the fairness-aware regularization term controls the recommendations being toward popular items. %Also, Steck in \cite{harald2011} examined the trade-off between degrading accuracy for improving long-tail coverage. By conducting user study, they observed that adding a small bias toward long-tail items leads to better feedback from users.   

Jannach et al., \cite{Jannach2015} conducted a comprehensive set of analysis on popularity bias of several recommendation algorithms. They analyzed recommended items by different recommendation algorithms in terms of their average ratings and their popularity. While it is very dependent to the characteristics of the data sets, they found that some algorithms (e.g., \algname{SlopeOne}, KNN techniques, and ALS-variant of factorization models) focus mostly on high-rated items which bias them toward a small sets of items (low coverage). Also, they found that some algorithms (e.g., ALS-variants of factorization model) tend to recommend popular items, while some other algorithms (e.g., \algname{UserKNN} and \algname{SlopeOne}) tend to recommend less-popular items. 

%Ge et. al., \cite{ge2010} analyzed the trade-off between coverage and serendipity of the recommendations and proposed a new way of measuring these metrics. In their proposed measurements, the usefulness of an item is incorporated to measure the coverage and serendipity in a way to enhance the quality of recommendations. 

%\subsection{Multi-Stakeholder Recommendations}

Multi-stakeholder recommender systems simultaneously take into account the fairness of all stakeholders or entities in a multi-sided platform. The main goal of multi-stakeholder recommendations is maximizing the fairness of all stakeholders. Consumers and providers are the major stakeholders in most multi-sided platforms \cite{burke2016,himan2019}.

Surer et al. in \cite{surer2018} proposed a multi-stakeholder optimization model that works as a post-processing approach for standard recommendation algorithms. In this model, a set of constraints for providers are considered when generating recommendation lists for end users. Also, Liu and Burke in \cite{liu2018} proposed a fairness-aware re-ranking approach that iteratively balances the ranking quality and provider fairness. In this post-processing approach, users' tolerance for diversity list is also considered to find trade-off between accuracy and provider fairness. 

\section{Fairness Metrics}

In this paper, we compare the performance of state-of-the-art recommendation algorithms in terms of bias disparity in recommended lists. We also consider ranking quality and item coverage of recommendation algorithms as two important additional metrics.

%Bias disparity measures how recommendation outputs deviate from groups' interest across item categories. Low bias disparity indicates that the algorithm does not amplify the original bias in the preference data.

%In this paper, we address \textit{consumer-side} and \textit{provider-side fairness} in recommendation outputs. In consumer-side fairness, we are interested to see how equally groups' interest across item categories are taken into account in recommendation lists. For example, consider the situation that female and male showed the same interest to \textit{Action} movies, but in recommendation results, female received much lower movies in \textit{Action} genre. This is a clear discrimination in recommendation output. On the other hand, in provider-side fairness, we are interested to see how recommendation algorithms equally show items belong to different providers in recommended lists. This means that all providers must have equal chance to be shown to users for satisfying provider-side fairness.  

We use two metrics to measure changes in bias for groups of users given item categories: \textit{bias disparity} and \textit{average disparity}. 

Bias disparity measures how much an individual's recommendation list deviates from his or her original preferences in the training set \cite{virginia2018}. Given a group of users, $G$, and an item category, $C$, bias disparity is defined as follow:

\begin{equation} \label{eq:bd}
BD(G,C)=\frac{B_R(G,C)-B_T(G,C)}{B_T(G,C)}
\end{equation}
where $B_T$ ($B_R$) is the \textit{bias} value of group $G$ on category $C$ in training data (recommendation list). $B_T$ is defined by:

\begin{equation} \label{eq:b}
B_T(G,C)=\frac{PR_T(G,C)}{P(C)}
\end{equation}

where $P(C)$ is the fraction of item category $C$ in the dataset defined as $|C|\textbackslash|m|$. $PR_T$ is the preference ratio of group $G$ on category $C$ calculated as:

\begin{equation} \label{eq:pr}
PR_T(G,C)=\frac{\sum_{u \in G} \sum_{i \in C} T(u,i)}{\sum_{u \in G} \sum_{i \in I} T(u,i)}
\end{equation}

where $T$ is the binarized user-item matrix. If user $u$ has rated item $i$, then $T(u,i)=1$, otherwise $T(u,i)=0$.

The \textit{bias} value of group $G$ on category $C$ in the recommendation list, $B_R$, is defined similarly. 

On the other hand, \textit{average disparity} measures how much preference disparity between training data and recommendation list for one group of users (e.g., unprotected groups) is different from that for another group of users (e.g., protected group). Inspired by \textit{value unfairness} metric proposed by Yao and Huang \cite{yao2017}, we introduce the average disparity as:

\begin{equation} \label{eq:fairness}
\begin{aligned}
\overline{disparity}=\frac{1}{|C|} \sum_{i=0}^{|C|}|(N_R(G_U,C_i)-N_T(G_U,C_i)) \\
-(N_R(G_P,C_i)-N_T(G_P,C_i))|
\end{aligned}
\end{equation}

where $G_U$ and $G_P$ are unprotected and protected groups, respectively. $N_R(G,C)$ and $N_T(G,C)$ return number of items from category $C$ in recommendation lists and training data, respectively, that are rated by users in group $G$. 

%Also, to measure provider-side fairness, we calculate the item coverage of recommended lists. 
As part of our analysis, we also measure item coverage of recommended lists which is an important consideration in provider-side fairness. Given the whole set of items in the system, $I$, and whole recommendation lists for all users, $R_{all}$, item coverage measures what percentage of items in the system appeared in recommendation lists and can be calculated as:

\begin{equation}
coverage=100.\frac{|\{i, i \in (R_{all} \cap I)\}|}{|I|} 
\end{equation} 

\section{Experiments}

\subsection{Experimental setup}

For comparing the effects of recommendation algorithms on bias and on item coverage, we performed an extensive experiments on state-of-the-art recommendation algorithms. Experiments are performed on model-based, neighborhood-based, and trust-aware recommendation algorithms. 

Our experiments on neighborhood-based recommendation algorithms include user-based collaborative filtering (\algname{UserKNN}) \cite{Resnick:1994a} and item-based collaborative filtering (\algname{ItemKNN}) \cite{sarwar2001}. Also, our experiments on model-based recommendation algorithms include biased matrix factorization (\algname{BiasedMF}) \cite{Koren:2009a}, combined explicit and implicit model (\algname{SVD++}) \cite{Koren:2008a}, list-wise matrix factorization (\algname{ListRankMF}) \cite{shi2010}, and the sparse linear method (\algname{SLIM}) \cite{Ning2011}. Finally, our experiments on trust-aware recommendation algorithms include trust-aware neighborhood model (\algname{TrustKNN}) \cite{Massa:2007a}, trust-based singular value decomposition (\algname{TrustSVD}) \cite{Guibing2015}, social regularization-based method (\algname{SoReg}) \cite{ma2011}, trust-based matrix factorization (\algname{TrustMF}) \cite{Bo:2017}, and social matrix factorization (\algname{SocialMF}) \cite{jamali2010}. Besides above well-known recommendation algorithms, we also performed experiments on two naive algorithms: random and most popular. 

For sensitivity analysis, we performed extensive experiments with different parameter configurations for each algorithm. Table \ref{tab:params} shows the parameter configurations we used for our experiments. %Among all those results, we report the best performing result for each algorithm.

\input{table-params.tex}

We performed 5-fold cross validation, and in the test condition, generated recommendation lists of size 10 for each user. Then, we evaluated nDCG, item coverage, bias disparity, and average disparity at list size 10. Results were averaged over all users and then over all folds. We used \textit{librec-auto} and LibRec 2.0 for all experiments \cite{mansoury2018automating,Guo2015}.

\begin{figure*}[t]
  %\centering
  \begin{subfigure}[b]{1\textwidth}
        \includegraphics[width=1\textwidth]{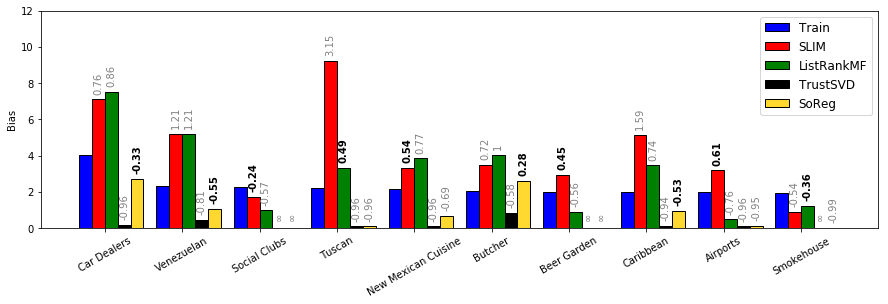}
        \caption{Male} \label{fig:mf:male}
  \end{subfigure}
  \begin{subfigure}[b]{1\textwidth}
        \includegraphics[width=\textwidth]{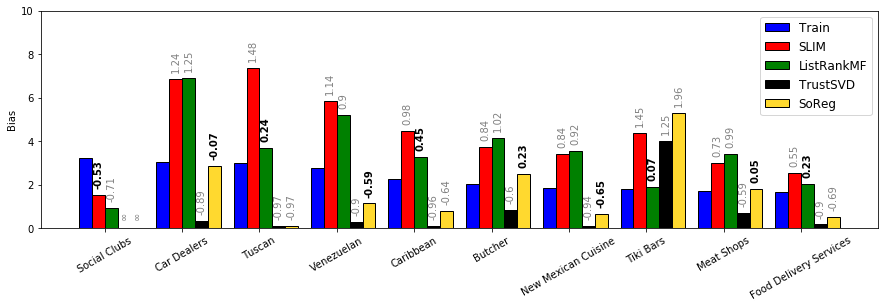}
        \caption{Female} \label{fig:mf:female}
  \end{subfigure}%
  \caption{Bias disparity for model-based recommendation algorithms. The x-axis is the top 10 most preferred categories for male and female on training data and y-axis is bias value computed by equation \ref{eq:b}. The numbers on each bar shows the bias disparity computed by equation \ref{eq:bd}. Numbers in bold show the lowest bias disparity for each category.} \label{fig:mf}
\end{figure*}

\begin{figure*}[t]
  %\centering
  \begin{subfigure}[b]{1\textwidth}
        \includegraphics[width=1\textwidth]{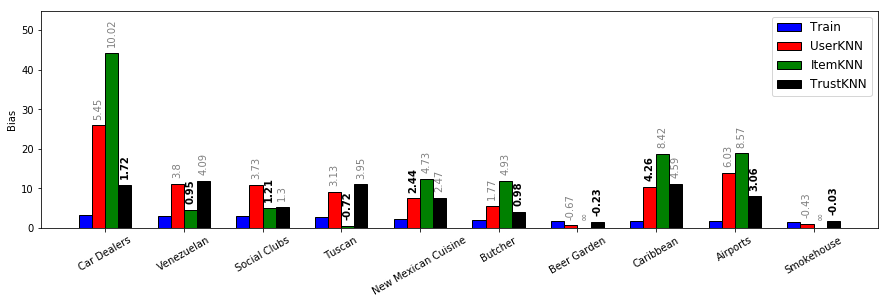}
        \caption{Male} \label{fig:knn:male}
  \end{subfigure}
  \begin{subfigure}[b]{1\textwidth}
        \includegraphics[width=\textwidth]{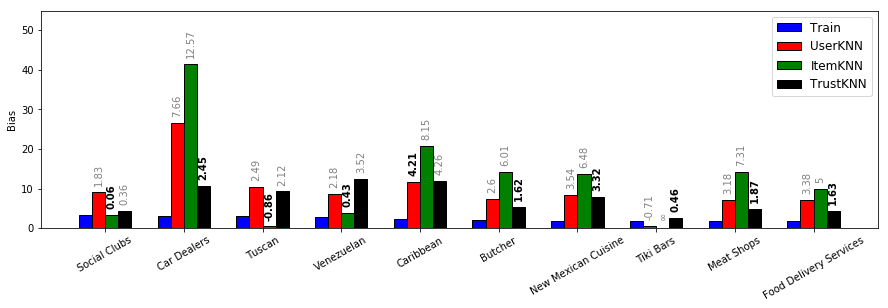}
        \caption{Female} \label{fig:knn:female}
  \end{subfigure}%
  \caption{Bias disparity for memory-based recommendation algorithms. The x-axis is the top 10 most preferred categories for male and female on training data and y-axis is bias value computed by equation \ref{eq:b}. The numbers on each bar shows the bias disparity computed by equation \ref{eq:bd}. Numbers in bold show the lowest bias disparity for each category.} \label{fig:knn}
\end{figure*}

\begin{figure*}[htp]
  \centering
  \begin{subfigure}[b]{0.438\textwidth}
        \includegraphics[width=\textwidth]{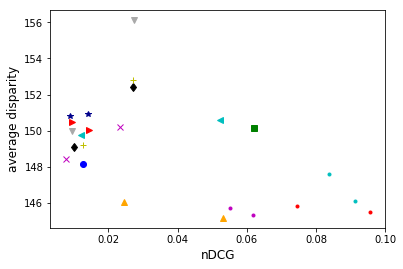}
        \caption{nDCG vs. average disparity} \label{fig:overall:unfairness}
  \end{subfigure}
\begin{subfigure}[b]{0.552\textwidth}
\includegraphics[width=\textwidth]{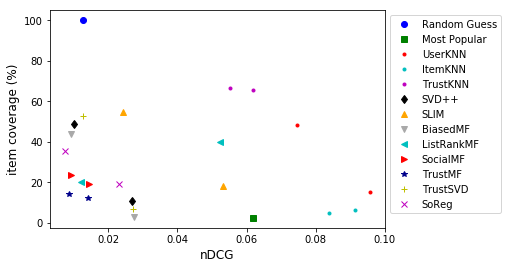}
\caption{nDCG vs. item coverage} \label{fig:overall:coverage}
\end{subfigure}%
\caption{Comparison of recommendation algorithms by ranking quality and item coverage/average disparity.} \label{fig:overall}
\end{figure*}

\subsection{Yelp dataset}

For our experiments, we use a subset of Yelp dataset from round 12 of Yelp Challenge\footnote{https://www.yelp.com/dataset}. In this sample, each user has rated at least 40 businesses and each business is rated by at least 40 users. Thus, there are 1,355 users who provided 100,409 ratings on 1,272 businesses. The range of ratings is 1 (not preferred) to 5 (preferred). The density of rating matrix is 5.826.

This Yelp dataset also has information about users friendship. Each user has selected a set of other users as her friends. We interpret this relationships as a trust network. When user $A$ selects user $B$ as a friend, it means that user $A$ trusts user $B$ with respect to the corresponding domain or category. In this dataset, 919 users have expressed their trustworthiness to 1,172 users and there are 26,453 trust relationships between users. With regard to the number of users, the density of trust matrix is 2.456.

In order to evaluate the recommendation outputs in terms of bias disparity and average disparity, specific information about users and items is needed. First, we need to define users group based on users demographic information and item category based on item contents. In Yelp dataset, there is no useful information about user to define users' group. To overcome this issue, we prepared the dataset by extracting users' gender from users' name. To do this, we use an existing online tool\footnote{https://gender-api.com} to extract users' gender. In this tool, for each user name as input, it will return the predicted gender, number of samples used for prediction, and prediction accuracy. Hence, it enables us to increase the reliability of extracted genders by taking outputs with high accuracy and fair amount of samples.

Moreover, information about items' category is provided in the dataset. Each business in Yelp dataset is assigned multiple relevant categories.

Overall, the prepared dataset has four separate sets:

\begin{enumerate}[\hspace{0.4cm}1.]
  \item The rating data that each user provided to businesses.
  \item Explicit trust data that each user has selected trusted (friends) users.
  \item Users information that consists of users' gender.
  \item Items category that consists of several category for each business.
\end{enumerate}

By using this dataset, we define the set $G=<male, female>$ and set $C$ as categories assigned to each business. The dataset is available at \href{https://github.com/masoudmansoury/yelp\_core40}{https://github.com/masoudmansoury/yelp\_core40}.

\subsection{Experimental results}

In this section, we compare the performance of recommendation algorithms across the different metrics discussed earlier. First, we show the bias disparity of recommendations results on top 10 most preferred item categories. Second, we show average disparity for each algorithm on all categories. For sensible comparison, we also take into account the ranking quality and item coverage.

\subsubsection{Bias disparity}

Results on model-based recommendation algorithms on top 10 most preferred item categories for male and female are shown in Figure \ref{fig:mf}. Figure \ref{fig:mf:male} shows the bias disparity for male individuals and Figure \ref{fig:mf:female} shows the bias disparity for female individuals. Since there is always a trade-off between accuracy and non-accuracy metrics (e.g., nDCG vs. fairness), for comparison, the fairness analysis is conducted on recommendation outputs that give the same nDCG (highest possible) for all recommendation algorithms. For model-based recommendation algorithms, the nDCG value is set to $0.023\pm0.001$. %We also consider the ranking quality of random and most popular item recommendations as baseline when choosing the nDCG value. 
This setting guarantees that the fairness of recommendation algorithms is compared in same condition for all algorithms.

As it is shown in Figure \ref{fig:mf}, in most cases, \algname{SoReg} provides lower bias disparity on top 10 most preferred categories for male and female groups. For males in Figure \ref{fig:mf:male}, \algname{SoReg} and \algname{SLIM} generated more stable outputs compared to other algorithms with the lowest bias disparity in 40\% cases. On the other hand, for female, \algname{SoReg} and \algname{ListRankMF} generated recommendations with the lowest bias disparity of 50\% and 40\% cases, respectively, when compared to other recommendation algorithms.

In Figure \ref{fig:mf}, we did not report the results for \algname{BiasedMF}, \algname{SVD++}, \algname{SocialMF}, \algname{TrustMF}, and random and most popular item recommendations because these algorithms either did not recommend any items from top 10 most preferred categories, or their ranking quality was lower than specified value for other algorithms.  

%\algname{TrustSVD} is another trust-based recommendation algorithm reported in \ref{fig:mf}. Unlike \algname{SoReg}, \algname{TrustSVD} is unable to generate fair recommendations and its bias disparity value is worse than other recommendation algorithms. One possible reason for this bad performance can be the sparsity of the trust data. Also, we performed experiments on \algname{SVD++}, \algname{BiasedMF}, and \algname{TrustMF}, but these algorithms did not generate any recommendation for those top 10 categories. 

Results on neighborhood-based recommendation algorithms for male and female groups are shown in Figure \ref{fig:knn}. The nDCG values for neighborhood algorithms are all set to $0.074\pm0.01$. Figure \ref{fig:knn:male} shows the bias disparity of neighborhood models for male. \algname{TrustKNN} generated more stable recommendations compared to other algorithms with 50\% top 10 most categories. Also, for other categories, its output is very close to the best one. Moreover, a better output in terms of bias disparity can be observed in Figure \ref{fig:knn:female} for female. On 60\% of top 10 most preferred categories, \algname{TrustKNN} worked better that other neighborhood algorithms. 

\subsubsection{Average disparity}

Figure \ref{fig:overall} compares the performance of recommendation algorithms with respect to two criteria: 1) how accurately recommendation algorithms generate stable (i.e. low disparity) recommendations for unprotected and protected groups, 2) how accurately recommendation algorithms are able to equally recommend the items belonging to all providers when generating recommendations (provider-side fairness). 

%The results in figure \ref{fig:overall} are based on ranking quality of recommendation algorithms. 
For all experiments that we performed with different hyperparameters, the best and worst nDCG for each algorithm are reported in Figure \ref{fig:overall}.

Random guess algorithm is a naive approach that randomly recommends a list of items to each user. Although this algorithm has low accuracy, it has the highest item coverage and lower average disparity compared to other recommendation algorithms. This algorithm does not take any preferences into account and unlikely to provide good results for any user. Also, most popular item recommendation is another naive, non-personalized, algorithm that only recommends items with the highest number of ratings to each user. Although it has high ranking quality and average disparity similar to model-based recommendation algorithms, it has the lowest item coverage. These algorithms provide baselines that other algorithms should be expected to beat.    

For neighborhood models, \algname{TrustKNN} showed better performance. Although it has lower ranking quality than \algname{UserKNN} and \algname{ItemKNN}, it has significantly better item coverage and average disparity. One possible reason for low nDCG of \algname{TrustKNN} can be high sparsity of trust matrix. Using a propagation model for reducing the sparsity of trust matrix may increase the ranking quality of \algname{TrustKNN}. Overall, neighborhood algorithms worked better than model-based algorithms in terms of all metrics. This is due to the fact that the rating data for these experiments is very dense and all users are heavy raters. 

For model-based algorithms, \algname{SLIM} shows better performance compared to other algorithms. From Figure \ref{fig:overall:unfairness}, while showing high nDCG, it has the lowest average disparity and in terms of item coverage, it has comparable coverage to other model-based algorithms. This result is also consistent with the definition of \algname{SLIM} algorithm which is an extension of \algname{ItemKNN} and analogous to neighborhood algorithms, it showed significant performance.

In addition, \algname{ListRankMF} is another model-based algorithm that, although having high accuracy and item coverage, has average disparity is as high as other algorithms. Also, for model-based trust-aware recommendation algorithms, although \algname{SoReg} showed significant reduction in bias disparity on the top 10 most preferred categories, it did not improve the average disparity on all categories.

\section{Conclusion}

In this paper, we examined the effectiveness of recommendation algorithms in generating outputs with lower bias disparity for different groups of users across item categories. We measured the performance of recommendation algorithms in terms of bias disparity on top 10 most preferred item categories, average disparity, ranking quality, and item coverage. A comprehensive sets of experiments showed that neighborhood models work significantly better than other algorithms, particularly trust-aware neighborhood model that outperformed other algorithms. Also, we observed that in most cases, having additional information along with rating data can enhance the performance of recommender systems. 

For future work, we would like to investigate individual fairness by considering the performance of recommendation algorithms in capturing individual users' interest across different item categories. Also, we are interested to repeat the experiments in this paper on another sample of Yelp dataset with sparser rating data and denser trust data to see how recommendation algorithms are able to control bias disparity. 

\bibliographystyle{ACM-Reference-Format}
\bibliography{sample-base}

\end{document}

%% file: table-params.tex
\captionsetup[table]{skip=4pt}
\begin{table}[t!]
\small
\centering
\captionof{table}{Parameter configuration} \label{tab:params}
\begin{tabular}{l|l}
\toprule
 parameter & values \\
 \midrule
 \#neighbors & \{10,20,30,40,50,70,100,200\} \\
 shrinkage & \{10,30,50,100,200\} \\
 similarity & \{pcc,cos\} \\
 user regularization & \{0.0001,0.001,0.005,0.01\} \\
 item regularization & \{0.0001,0.001,0.005,0.01\} \\
 bias regularization & \{0.0001,0.001,0.005,0.01\} \\
 implicit regularization & \{0.0001,0.001,0.005,0.01\} \\
 learning rate & \{0.0001,0.001,0.005,0.01\} \\
 \#iterations & \{10,30,50,100\} \\
 \#factors & \{10,30,50,100,150,200,300\} \\
 $\ell_{1}$-norm & \{0.005,0.05,0.5,2,5\} \\
 $\ell_{2}$-norm & \{0.005,0.05,0.5,2,5\} \\

\bottomrule
\end{tabular}
\end{table}